# ERRORS IN A POLYNOMIAL-TIME ALGORITHM FOR 3-SAT

## Ritesh S. Vispute

There are errors in the algorithm proposed by Narendra Chaudhari [2] purporting to solve the 3⁻sat problem in polynomial time. The present paper presents an instance for which the algorithm outputs erroneous results.

**Keywords:**

3-sat, satisfiable, unsatisfiable

**Introduction:**

In his paper [1], Narendra Chaudhari describes an algorithm to find a solution for the 3-sat problem in polynomial time. In a second paper [2], he improves on his previous algorithm. The following paper describes two instances of the 3-sat problem on which the improved algorithm fails to give the expected output.

The following algorithm was also presented in Cybernetics and Intelligent Systems (CIS), 2010 IEEE Conference.

**Details:**

In both instances, the variables are a1, a2, … a9, total 9 variables.

**Instance 1:** Clauses in conjunctive normal form (CNF):

(a1 a2 a3)  (a1 -a2 -a3) (-a1 a2 -a3) (-a1 -a2 a3) (-a4 -a5 a6) (-a4 a5 -a6) (a4 a5 a6) (a4 -a5 -a6) (a7 a8 a9) (a7 -a8 -a9) (-a7 a8 -a9) (-a7 -a8 a9) (-a1 -a4 -a7) (-a1 a4 a7) (a1 a4 -a7) (a1 -a4 a7) (-a2 -a5 -a8) (-a2 a5 a8) (a2 -a5 a8) (a2 a5 -a8) (-a3 a6 a9) (-a3 -a6 -a9) (a3 a6 -a9) (a3 -a6 a9)

**Instance 2:** Clauses in conjunctive normal form (CNF):

(a1 a2 a3)  (a1 -a2 -a3) (-a1 a2 -a3) (-a1 -a2 a3) (a4 a5 a6) (a4 -a5-a6) (-a4 a5 -a6) (-a4 -a5 a6) (a7 a8 a9) (a7 -a8 -a9) (-a7 a8 -a9) (-a7 -a8 a9) (a1 a4 a7) (a1 -a4 -a7) (-a1 a4 -a7) (-a1 -a4 a7) (a2 a5 a8) (a2 -a5 -a8) (-a2 a5 -a8) (-a2 -a5 a8) (a3 a6 a9) (a3 -a6 -a9) (-a3 a6 -a9) (-a3 -a6 a9)

The W-closure $W_{(x,y)}$ has been defined in [1] for literal pairs (x, y). For both these instances, while computing $W_{(x,y)}$ where x ≠ y, the R-working set of restricted literals [1] includes the variables but not their negations. According to the algorithm, in this situation $W_{(x,x)}$ for any x does not get updated, i.e., $W_{(x,x)}$ remains initialized to {x} throughout the algorithm.

Furthermore for both instances, in substep 2 of step IV the T-collection of top-level restricted literals never gets updated and always remains empty as initialized at the start of step III.

Again for both instances, step 3 and step 4 of the algorithm never gets executed.

As a result, for both instances, the algorithm exits with satisfiable-flag := True and T := { }.

**Instance 1:** is unsatisfiable but the algorithm erroneously outputs 'satisfiable'.

The claim that this instance is unsatisfiable has been verified with the publicly available algorithm March_eq [3]. Thus the algorithm yields a "false positive" result.

**Instance 2:** the algorithm outputs 'satisfiable', but no satisfying truth value assignments are offered to verify the claim. In this sense the algorithm is incomplete.

It can be readily seen that this instance is satisfiable when all of the variables a1 to a9 are set to 1.

**Conclusion:**

The error in instance 1 proves that the algorithm yields a "false positive" result with respect to some of the instances as shown above.

**References:**

1. Narendra S. Chaudhari, "Computationally Hard Problems: 3-sat and Its Polynomial Satisfiablity", Journal Of Indian Academy Of Mathematics, Vol. 31, No. 2 (2009). http://dcis.uohyd.ernet.in/~wankarcs/index_files/pdf/Vol-31-No-2-2009-pp407-444-scanned-copy.pdf
2. Narendra S. Chaudhari, "Improved Polynomial Algorithm For 3-sat", Journal Of Indian Academy Of Mathematics, Vol. 32, No. 1 (2010)
3. March_eq, http://www.st.ewi.tudelft.nl/sat/march_eq.htm